\documentclass[jcp,amsmath,amssymb,showkeys,floatfix,
reprint
]{revtex4-1}

\usepackage{bibentry}
\usepackage{appendix}
\usepackage[framemethod=tikz]{mdframed}
\mdfsetup{
	outerlinewidth=1pt,
	linecolor=white!90!black,
	innertopmargin=6pt,
	innerbottommargin=6pt,
	leftmargin=2pt,
	rightmargin=2pt
}
\usepackage[utf8]{inputenc}
\usepackage{float}

\usepackage{color}

\usepackage{graphicx}
\usepackage{dcolumn}
\usepackage{bm}
\usepackage{xfrac}
\usepackage{colortbl}

\usepackage{natbib}
\usepackage{bibentry}
\usepackage{float}
\usepackage{sidecap}
\usepackage[section]{placeins}

\usepackage[caption=false]{subfig}

\usepackage{amsmath}
\usepackage{amssymb}
\usepackage{physics}
\usepackage{marvosym}
\usepackage{wasysym}
\usepackage{pifont}
\usepackage{wrapfig}
\usepackage{xifthen}
\usepackage{enumerate}
\usepackage{cancel}
\usepackage{MnSymbol}
\usepackage{mathtools}
\usepackage{relsize}

\usepackage[normalem]{ulem}

\begin{document}

\title{Quantum Alternating Operator Ansatz (QAOA) Phase Diagrams and Applications for Quantum Chemistry}

\author{Vladimir Kremenetski }
\affiliation{Quantum Artificial Intelligence Laboratory (QuAIL), Exploration Technology Directorate, NASA Ames Research Center, Moffett Field, CA 94035, USA}
\affiliation{USRA Research Institute for Advanced Computer Science, Mountain View, California 94043, USA}
\author{Tad Hogg}
\affiliation{Quantum Artificial Intelligence Lab. (QuAIL), Exploration Technology Directorate, NASA Ames Research Center, Moffett Field, CA 94035, USA}
\author{Stuart Hadfield}
\affiliation{Quantum Artificial Intelligence Lab. (QuAIL), Exploration Technology Directorate, NASA Ames Research Center, Moffett Field, CA 94035, USA}
\affiliation{USRA Research Institute for Advanced Computer Science, Mountain View, CA 94043, USA}
\author{Stephen J. Cotton}
\affiliation{Quantum Artificial Intelligence Laboratory (QuAIL), Exploration Technology Directorate, NASA Ames Research Center, Moffett Field, CA 94035, USA}
\affiliation{KBR, 601 Jefferson St., Houston, TX 77002}

\author{ Norm M. Tubman}
\email{norman.m.tubman@nasa.gov}
\affiliation{Quantum Artificial Intelligence Laboratory (QuAIL), Exploration Technology Directorate, NASA Ames Research Center, Moffett Field, CA 94035, USA}
\date{\today}
\begin{abstract}

We adapt the Quantum Alternating Operator Ansatz (QAOA) to find ground states in quantum chemistry problems and empirically evaluate our protocol on several molecules. These simulations show robust qualitative behavior for QAOA as a function of the number of steps and size of the parameters. This includes cases where non-adiabatic schedules achieve state preparation accuracy close to that of the adiabatic limit while using relatively low quantum circuit depth. Simulations also show these behaviors in QAOA applied combinatorial search, providing evidence of the generality of these behaviors.
\end{abstract}

\maketitle

One of the most promising applications of quantum computing is simulating physical systems, for which a fundamental task is computing the energies of ground and excited states~\cite{yung2012, bravyi2020}. Algorithms for this purpose include Hamiltonian simulation~\cite{lloyd1996universal,berry2015simulating} and quantum phase estimation~\cite{cleve1998quantum}, 
as well as variational approaches and related Lanczos type methods such as
variational quantum phase estimation~\cite{klymko2021real}.

Many of these methods rely on being able to prepare a sufficiently good approximate ground or excited state.
Adiabatic state preparation (ASP)~\cite{aspuru2005simulated,crosson2020prospects,hauke2020perspectives,brady2020optimal,krem2021} is one approach, which entails slowly evolving a known ground state of a simple system to the ground state of a system of interest. The system remains essentially in the ground state of the evolving Hamiltonian provided that the evolution time is large compared to the inverse square of the minimum gap between the first two energies \cite{albash2018}. In particular, this means the system will be in the target ground state at the end of the evolution.  

A popular alternative to ASP, especially on near-term devices, is the variational quantum eigensolver (VQE)~\cite{peruzzo2014,mcclean2016}. This is an iterative process where, at each step, a sequence of parameterized quantum gates is applied to an initial state and the expected energy of the resulting state is measured. A classical algorithm uses the measurement results to adjust the gate parameters in an attempt to minimize the energy for the next measurement~\cite{peruzzo2014}. This quantum-classical hybrid method can work on NISQ devices~\cite{mcclean2016}, though finding good parameters can be prohibitively costly as the number of parameters increases with circuit depth~\cite{mcclean2016, wierichs2020, mcclean2018}.

In this paper we investigate a modified version of a 
distinct but related 
class of 
parameterized quantum circuits for finding molecular ground states and energies, the quantum approximate optimization algorithm (QAOA)~\cite{farhi2014,hogg2000,harrigan2021} and its generalization to the quantum alternating operator ansatz~\cite{hadfield2019quantum} that incorporates broader classes of operators and states.   QAOA has been demonstrated to outperform adiabatic techniques in certain contexts~\cite{zhou2020}.
Research on QAOA has largely focused on combinatorial optimization problems encoded as a cost Hamiltonian $H_C$ whose ground state encodes the optimal problem solution. 
The level-$p$ QAOA algorithm uses $H_C$ and a ``mixer'' Hamiltonian $H_B$ as follows:
An easy-to-prepare initial state, for example the ground state of $H_B$, is prepared. 
For the  originally proposed transverse-field mixer,  
the initial state is $\ket{\psi_0} = \ket{+}^{\otimes N}$, an equally weighted superposition over the computational basis. 
An alternating sequence of of length $2p$ of parameterized unitary operators is then applied to 
yield the state
\begin{equation} \label{eq:QAOAstate}
    \ket{\psi_p(\vec{\gamma},\vec{\beta})} = e^{-i\beta_pH_B}e^{-i\gamma_pH_C}...e^{-i\beta_1H_B}e^{-i\gamma_1H_C}\ket{\psi_0}.
\end{equation}
The state can then be measured to estimate quantities such as expectation values, and reprepared with updated parameters. 
For sufficiently large $p$ there exists a set of parameters that yields an arbitrarily good approximation to the ground state~\cite{farhi2014}.

For applications in quantum chemistry, we evaluate QAOA performance in terms of the squared overlap of the final QAOA state with the ground state $\ket{\phi_{\text{gs}}}$: 
\begin{equation} \label{eq:squaredOverlap}
    \mbox{squared overlap} = \left| \bra{\phi_{\text{gs}} }\ket{\psi_p(\vec{\gamma},\vec{\beta})} \right|^2
\end{equation}
If there are degenerate ground states, this measure generalizes to the sum of squared overlaps over all the ground states.
As explained below, this performance metric is well-suited for applications such as chemistry, but generally less suitable for classical optimization, where the QAOA state may give good approximate solutions with little or no support on the true ground state.

There are 
a number of ways to select suitable parameters $\vec{\gamma},\vec{\beta}$.
One approach is optimizing all of the parameters.
However, such optimization is expensive and may get stuck in local minima or encounter other 
bottlenecks such as exponentially small gradients~\cite{mcclean2016, wierichs2020, mcclean2018,ruslan2019,ruslan2020,ruslan2021}. Alternatively, one can define $\gamma_j$ and $\beta_j$ as functions of $j$ specified by just a few parameters, thereby drastically reducing the number of free parameters to be optimized.

Through this research we study the regime of QAOA with $p\gg 1$ for which relatively few results exist in the literature. This enables identification of generic performance behavior expressed through what we term the \textit{QAOA phase diagram}, which serves multiple purposes including a visualization of the performance of QAOA in the large p limit. For the molecules that we consider, we empirically demonstrate that simple QAOA parameter schedules may offer significant improvement over 
algorithms in 
the adiabatic regime. 

\textbf{QAOA for chemistry:}  We consider determining the ground state energy of a molecule with $n$ electrons, whose Hamiltonian is
$H_e = \sum_{i<j}h_{ij} a^\dagger_i a_j
+ \sum_{ijk\ell}h_{ijk\ell} a^\dagger_i a^\dagger_j a_ka_\ell
\,+~\mbox{h.c.}$ where the $a_i$ $(a_i^\dagger)$
are Fermionic annihilation (creation) operators in a molecular orbital basis, and h.c. denotes the Hermitian conjugate. 
In this \textit{second-quantized} formulation the electronic Hamiltonian is  quartic in the creation and annihilation operators for each orbital~\cite{szabo2012modern}. 
We also use the Hartree-Fock Hamiltonian, $H_\text{HF}$, which is diagonal in the basis of Slater determinants and defined by the energies of the single-determinant Hartree-Fock approximate ground state and its excitations.

We truncate the Hilbert space and $H_e$ to $N>n$ single-particle spin-orbitals, which account for both position and spin degrees of freedom. 
We consider the Jordan-Wigner representation where 
Fermion occupation number states are mapped to $N$-qubit computational basis states $\ket{n_1n_2\dots n_N}$, which gives
$ H_e = \sum_{j=1}^m H_j = \sum_{j=1}^m c_j \sigma_j,$ 
where $\sigma_j$ are strings of single-qubit Pauli operators. The number of terms is 
$m\sim N^4$, though this may be reduced substantially in particular basis set representations.

\paragraph{Differences from optimization setting}
We present a modified version of QAOA 
adapted to chemistry applications.
This requires a different strategy than for combinatorial optimization, where QAOA aims to find a 
low energy 
solution (bitstring) 
that is obtained by measuring the QAOA state. 
As quantum computers are not believed to be able to efficiently solve NP-hard problems, such approximate solutions are often the best we can hope to achieve, and the QAOA state may have little or no support on the ground states.   
This is a 
distinct goal from 
approximating the ground state in quantum chemistry, as generally the 
electronic Hamiltonian eigenstates are 
superpositions of computational basis states, 
for which we are looking to generate a wave function whose squared overlap with the true ground state exceeds 99\%.  
Hence in optimization, QAOA finds exact eigenvectors (approximate solutions) and energies (cost function evaluations) and returns the best solution found, whereas in QAOA for chemistry we seek a state with high ground state overlap that is necessarily close in expected energy.   

Chemistry applications differ from combinatorial optimization in other regards.   In particular,  
classical techniques provide many useful starting approximations in quantum chemistry,  which yields a well-motivated alternative to starting from an equal superposition of states in the Hilbert space~\cite{tubman2018}. 
In contrast, for optimization we often cannot do much better than random guessing in terms of initial state overlap with the optimal solution. 

Chemistry has at least two simulation goals of interest:  1) moderately accurate state preparation, that can be 
followed by 
further processing such as quantum 
phase estimation, and 
2) state preparation that attains a high accuracy state for which the 
estimated energy will be chemically accurate.
This study focuses on the second of these goals.

\paragraph{QAOA Ansatz}
Applying QAOA to chemistry requires defining suitable cost and mixer Hamiltonians in Eq.~\ref{eq:QAOAstate}. These must allow effective implementations, including creating the initial state as the ground state of the mixer.
We use the full Hamiltonian as the QAOA cost Hamiltonian, i.e., $H_C = H_e$.

Chemistry problems generally involve fixed particle number and fixed spin, and states that violate these quantities may be considered infeasible. 
Moreover, the Hartree-Fock state  
is an easily prepared feasible state~\cite{tubman2018} which typically has a relatively high squared overlap with the target state, i.e., the ground state of $H_e$. 
We take advantage of both these properties by using a mixer that remains in the subspace of feasible states and whose ground state is the Hartree-Fock state.
Specifically,
we use the Hartree-Fock state as the QAOA initial state and  
the Hartree-Fock Hamiltonian as the mixer: $H_B=H_\text{HF}$. 
This mixer preserves state feasibility (i.e., cannot change particle number), and 
is relatively cheap to implement 
as $H_\text{HF}$ involves
$\sim N^{2}$ operators, whereas $H_e$ 
generally requires 
$\sim N^{4}$
terms~\cite{szabo2012modern}.

A notable consequence of this choice of mixer and cost Hamiltonians is that it inverts the properties of the standard QAOA setup: our cost Hamiltonian is non-diagonal in the computational basis while our mixer is diagonal.  
This choice is directly inspired by adiabatic state preparation, though our approach generalizes beyond the adiabatic regime, as discussed below. 
A similar 
QAOA construction is briefly mentioned in \cite{bonet2018low}, although only in context of preserving symmetries. 

In contrast to our choice of mixer, the standard transverse-field X mixer does not keep the state vector in the feasible subspace. 
In addition, as described in Appendix~\ref{sec:appC}, while the XY mixer 
restricts evolution to 
the feasible subspace~\cite{hadfield2019quantum}, the Hartree-Fock state is not its ground state.
Thus these alternative mixers do not exploit the properties of the chemistry application as completely as the mixer we consider.

For simplicity and to reduce the dependence of our results on the quality of $\vec{\gamma},\vec{\beta}$ found through classical optimization, we focus on QAOA parameter schedules defined by linear ramps
\begin{equation}
    \gamma(f) = \Delta f  \quad\mbox{and}\quad  \beta(f) = \Delta(1-f)
    \label{e:gamma and beta}
\end{equation}
where $\Delta$ is a constant and $f$ ranges from 0 to 1.
The QAOA evaluation of Eq.~\ref{eq:QAOAstate} uses parameters sampled from these ramps: $\gamma_j = \gamma(f_j)$ and $\beta_j = \beta(f_j)$ where
\begin{equation}
    f_j = \frac{j}{p+1}
    \label{e:f}
\end{equation}
with $j=1,\ldots,p$.
Similar linear schedules 
are frequently used 
for QAOA 
in optimization settings~\cite{hogg2000,crooks2018performance,mbeng2019quantum,zhou2020,shaydulin2020classical,sack2021quantumannealing}.

We investigate the efficiency of state preparation of various $\Delta$ and $p$ for
several molecules. The product $\Delta p$ is roughly analogous to total annealing time of an analog version of this algorithm. 
 We use the same $\Delta$ for both $\gamma_j$ and $\beta_j$ due to the 
 commensurate energy scales of 
 the cost and mixer Hamiltonians.  These scales are similar because the mixer is the diagonal part of the cost Hamiltonian and the cost Hamiltonian is diagonal dominant for typical chemistry applications, and therefore determines the 
 spectrum of the mixer.

\paragraph{QAOA phase diagrams}
To 
show QAOA performance we introduce QAOA phase diagrams.
These diagrams show the squared overlap 
of Eq.~\ref{eq:squaredOverlap} 
as a function of $\Delta$ and $p$.  
QAOA phase diagrams are a useful pictorial representation of how QAOA performs among different problems and parameter choices.
Generalizing these diagrams to QAOA approaches that involve many variational parameters may not be as insightful.

\textbf{Time dynamics algorithm:} To test our approach on quantum chemistry applications and generate QAOA phase diagrams, we classically simulate systems of various sizes, including those that are generally larger than other studies with our recently introduced time evolution algorithm~\cite{krem2021}.  Doing a full configuration interaction (FCI) for the full cc-pVTZ orbital basis would require using up to $10^{19}$ determinants for the molecules we consider, which would be computationally infeasible. To circumvent this, we use a form of selective CI called adaptive sampling configuration interaction (ASCI)~\cite{tubman2016} to find the most important determinants for the first two eigenstates of our cost Hamiltonian, with the top $10^6$ determinants found for each. Due to the high squared overlap of our target state with our starting state, these determinants are the only ones necessary to capture the dynamics of our wavefunction as its squared overlap with the ground state changes with different parameter settings. The ASCI algorithm generates very accurate energies for electronic systems with strong many-body effects~\cite{tubman2016,tubman2020,mejuto2019} and provides an accurate description of time dynamics for wavefunctions close to the target ground state~\cite{krem2021}.

\begin{figure*}[htb]
\begin{center}
\includegraphics[width=\textwidth]{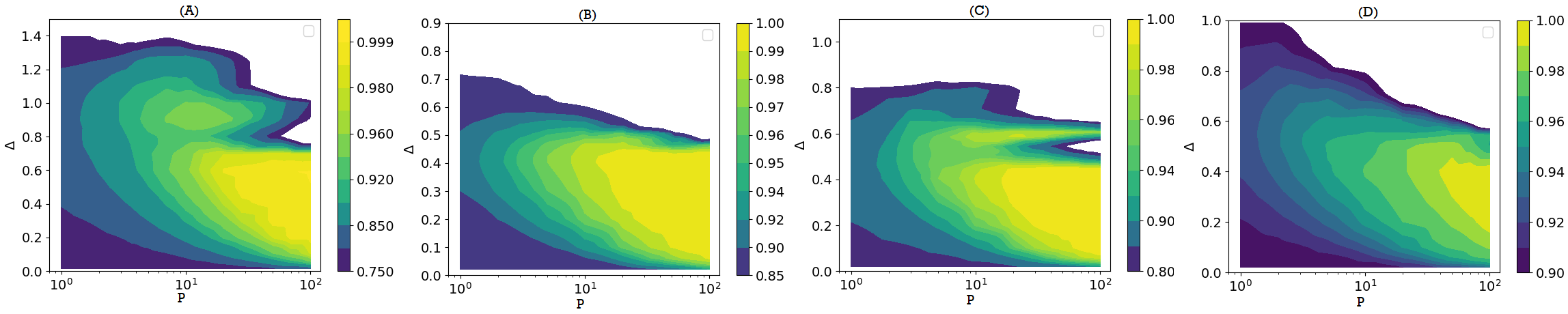}
\end{center}
\caption{In order: (A) P\textsubscript{2} with initial squared overlap 0.77, (B)  CO\textsubscript{2} with initial squared overlap 0.85, (C) Cl\textsubscript{2} with initial squared overlap 0.82, and (D) CH\textsubscript{2} with initial squared overlap 0.90. Colored regions are those where parameter settings result in an improved squared overlap. White regions above the colored areas represent parameter settings where squared overlap with target ground state is less than its initial value.   \label{fig7} }
\end{figure*}

\textbf{Molecular results:} We tested our modified QAOA method on four molecules at their equilibrium geometries: P\textsubscript{2}, CO\textsubscript{2}, Cl\textsubscript{2}, and CH\textsubscript{2}. All these molecules have unique ground states, and were selected from the G1 set (55 molecules)~\cite{headgordon1989}. Our choices are some of the harder molecules for our QAOA method because the squared overlap of their ground states with the Hartree-Fock state is lower than most molecules in this set. Specifically, the molecules we used have squared overlaps less than 0.9~\cite{tubman2018,krem2021}.

These test cases are somewhat large and would require over 100 logical qubits to test on quantum hardware.  For P\textsubscript{2} ($N=116$), Cl\textsubscript{2} ($N=116$), and CH\textsubscript{2} ($N=114$) the full cc-pVTZ orbital bases generated by Psi4~\cite{psi4} were used, with only the first sixty-four spatial orbitals of the cc-pVTZ basis used for CO\textsubscript{2} ($N=128$).
The values of $\Delta$ explored ranged from 0.01 to 6, with values of $p$ 
from 1 to 100. For all cases considered in this work, the initial wavefunction is the ground state of the Hartree-Fock Hamiltonian so the behaviors seen in our studies approach the adiabatic limit in certain cases.  For comparison, a recent study evaluating time evolution in the adiabatic limit with different schedules found that nearly linear interpolation between initial and target Hamiltonians outperformed all other monic polynomial interpolations~\cite{krem2021}.

Figure~\ref{fig7} shows the QAOA phase diagrams for our molecules. The white regions in these figures indicate parameter settings for which the squared overlap with the 
ground state decreases as a result of the protocol.  
We identify four main regions of interest in these diagrams: 
\setcounter{paragraph}{0}
\paragraph{Continuous adiabatic limit}
For very small $\Delta$ and very large $p$, 
QAOA 
becomes a close Trotter approximation to continuous evolution with a linear interpolation of the mixer and cost Hamiltonians. In particular, when 
\begin{equation}
    \Delta \rightarrow 0 \mbox{ and } \Delta p \rightarrow \infty
    \label{e:continuous adiabatic}
\end{equation}
this approximation matches the adiabatic limit of continuous evolution. This roughly corresponds to the bottom right corner of the diagrams, a region where, in accord with the adiabatic theorem, the squared overlap with the target ground state goes to 1 as $p$ increases. The corner itself is not completely yellow since the smallest values of $\Delta$ we evaluated were so small that $\Delta p$ is not large enough to approximate the limit of Eq.~\ref{e:continuous adiabatic}.   

\paragraph{Discrete adiabatic limit}
The other limit 
of particular interest is the \textit{discrete adiabatic limit}~\cite{hogg03} where 
\begin{equation}
    \mbox{$\Delta > 0$ is fixed and } p \rightarrow \infty
    \label{e:discrete adiabatic}
\end{equation}
This corresponds to the right edge of the phase diagrams.
In this case, the unitary QAOA operators for $j=1,\ldots,p$ change slowly as a function of $j$, so the evolving state vector ends up almost completely overlapping with the target by the end (yellow edge) up to a particular $\Delta$. For larger~$\Delta$, the overlap with the target ground state abruptly deteriorates (white edge).  For these larger $\Delta$, the state vector 
still evolves to an eigenstate of the cost Hamiltonian, but one that differs from the target.

\paragraph{Small $\Delta$}
The bottom half of each 
phase diagram shown in Figure~\ref{fig7} displays the expected monotonic improvement in QAOA performance with increasing $p$ for small values of $\Delta$.
Moreover, in this region, for fixed~$p$, QAOA performance improves as $\Delta$ increases, specifically for $\Delta$ increasing over a range with length of about 0.3 to 0.7.

The regime of small $\Delta$ and intermediate $p$ corresponds to a situation where the discrete adiabatic limit closely follows the continuous due to Trotter approximation.
Thus $p \Delta$ increases as $\Delta$ gets larger (for fixed $p$), corresponding to longer time for continuous adiabatic evolution. This, in turn, results in an improved overlap for fixed $p$ and increasing $\Delta$. In some cases, such behavior leads to choices of small $\Delta$ with 
intermediate 
values of $p$ giving even better overlaps than very small values of $\Delta$ but large $p$.
At large, but fixed, $p$, the limit $\Delta \rightarrow 0$ corresponds to a continuous evolution with a short time, which does not reach the adiabatic limit and gives worse performance than somewhat larger $\Delta$ at that value of $p$.

\paragraph{Large $\Delta$}
For fixed $p$, there is a sufficiently large value of $\Delta$, which we denote $\Delta_\text{crit}(p)$, above which QAOA performance starts to worsen, although not always monotonically. 
In particular, for the cases in Figure~\ref{fig7}, $\Delta_\text{crit}(p)$ is larger for small values of $p$ than for large values of $p$. 
Identifying, even roughly, the value of $\Delta_\text{crit}(p)$ 
for intermediate $p$ would allow maximizing the performance for such values of $p$, thereby giving high overlaps with significantly fewer steps than required to get similar overlap from approaching the continuous adiabatic limit.

\begin{figure}
    \centering
    \includegraphics[width=3.5in]{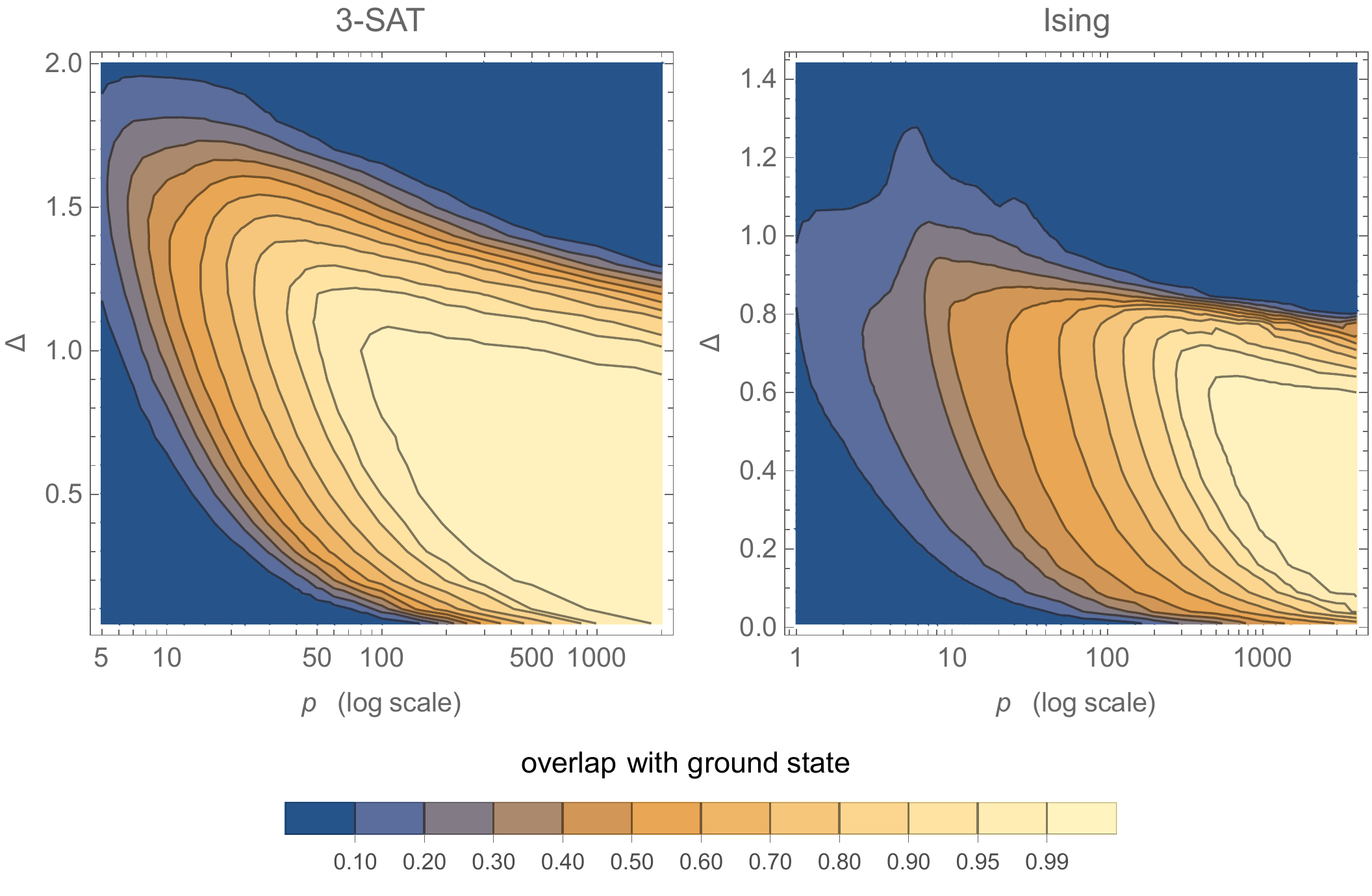}
    \caption{Squared overlap with ground state(s) after completing QAOA as a function of QAOA parameters $\Delta$ and $p$. On the left is a 20-variable instance of random 3-SAT with 80 clauses, and on the right is a 6-variable fully-connected Ising spin system. 
    }
    \label{fig:SAT-Ising}
\end{figure}

The qualitative behavior of QAOA shown in Fig.~\ref{fig7} is not specific to chemistry problems or our choice of mixer and initial state.
This behavior also occurs for combinatorial optimization problems using the standard $X$ mixer and uniform superposition initial state, as illustrated in Fig.~\ref{fig:SAT-Ising} for a randomly generated instances of 
satisfiability (SAT) and the Ising model; 
see Appendix~\ref{sec:appB} for details.


The phase diagrams identify novel features of QAOA that have not been discussed in the literature.  
For example, in the parameter regime with large $p$ and $\Delta$ (upper right white regions in Fig.~\ref{fig7}), the squared overlap tends toward zero as $p$ increases.
This behavior arises from a combination of the discrete adiabatic limit and how the QAOA operator eigenstates evolve during QAOA for different values of $\Delta$~\cite{hogg03}. 
  For large $\Delta$, the eigenvalues of the unitary QAOA operator ``wrap around'' the unit circle of the complex plane, becoming indistinguishable from eigenvalues for a unitary matrix generated by a Hamiltonian with a different ordering of eigenvalues.  Because of this, in the discrete adiabatic limit, following the instantaneous eigenvector of the QAOA operator that starts out close to the mixer Hamiltonian's ground state does not result in reaching the cost Hamiltonian's ground state by the end. Curiously, and very importantly, small scale model Hamiltonian simulations suggest following said eigenvector as $p\rightarrow \infty$ still results in reaching a single eigenstate, rather than a superposition of such states.  The eigenstate, however, is no longer the ground state of the cost Hamiltonian.  

In some cases, as $\Delta$ increases beyond $\Delta_\text{crit}(p)$
the change in performance is not monotonic. For instance, the phase diagrams for phosphorus and chlorine (Figs.~\ref{fig7}A and \ref{fig7}C, respectively) both have ridges with large squared overlap with the target extending to large $p$ at $\Delta$ somewhat larger than $\Delta_\text{crit}(p)$.

\textbf{Conclusion:} This paper shows how to adapt QAOA to chemistry Hamiltonians, and illustrates its performance 
through QAOA phase diagrams.  We 
used these diagrams to explore various QAOA parameter limits, including the continuous and discrete adiabatic regions.  For the application 
to chemistry 
we demonstrated that non-adiabatic evolution can be more efficient than the adiabatic limit for the schedules considered in this work. 
A more detailed comparison of the methods would require finding optimal adiabatic and optimal QAOA schedules. 
In the context of optimization there is a close connection between optimal QAOA parameters and smooth adiabatic schedules
~\cite{brady2020optimal,sack2021quantumannealing,venuti2021optimal,brady2021behavior}, showing conditions under which optimal schedules exist composed of ``bang-bang'' and ``annealing'' regimes, as well as the connection to counteradiabatic effects for suppressing excitations~\cite{wurtz2021counterdiabaticity}. 
Cases where there are relatively few free parameters may be especially amenable to generalizations of our phase diagram approach. 
A future direction is to 
formally extend these results 
and 
other proposed
approaches to analyzing QAOA (e.g., the  perturbation series analysis of  \cite{hadfield2021analytical}), to the chemistry setting.

Through our analysis of QAOA phase diagrams and chemistry Hamiltonians, we found 
common qualitative features 
of near-adiabatic QAOA schedules, as illustrated in our figures, and why non-adiabatic schedules can show improvements.  In particular, for the linear and nonlinear ramp schedules considered here, there is a region below a critical $\Delta$ for which 
schedules lead QAOA to converge to the ground state.  We demonstrate that for intermediate to small $p$, away from the adiabatic limit, increased convergence to the ground state can still be attained.  We show that these results extend beyond chemistry systems, including to combinatorial optimization, and so our phase diagrams help connect behaviors across multiple QAOA applications. 

\vspace{-3mm}
\section*{Acknowledgments:}
  We are grateful for support from NASA Ames Research Center.  
  This material is based upon work supported by the U.S. Department of Energy, Office of Science, National Quantum Information Science Research Centers, Superconducting Quantum Materials and Systems Center (SQMS) under the contract No. DE-AC02-07CH11359 and the Defense Advanced Research Projects Agency (DARPA) via IAA8839 annex 114.. 
  V.K., T.H., and S.H.  
  were supported by the NASA Academic Mission Services, Contract No. NNA16BD14C. 
  The work was started under DARPA funding. For the mature part of the work, NT and SH were supported by the SQMS funding and SC and TH by the DARPA funding.
  Calculations were performed as part of the XSEDE computational Project No. TG-MCA93S030 on  Bridges-2 at the Pittsburgh supercomputer center.

\bibliography{refs}

\appendix

\section{Nonlinear Schedules}

We use the linear ramps schedules defined in
Eq.~\ref{e:gamma and beta}.
However, nonlinear schedules can improve QAOA performance~\cite{zhou2020}. Additionally, nonlinear schedules improve the performance of adiabatic state preparation when they have slower changes in the neighborhood of small gaps~\cite{roland2002}. 
To check the generality of the behavior seen in the QAOA phase diagrams (rather than improve the schedule, since nonlinear schedules haven't shown much promise for molecular cases~\cite{krem2021}), we examined two nonlinear ramps. 

\begin{figure}[H]
    \centering
    \includegraphics[width=2.5 in]{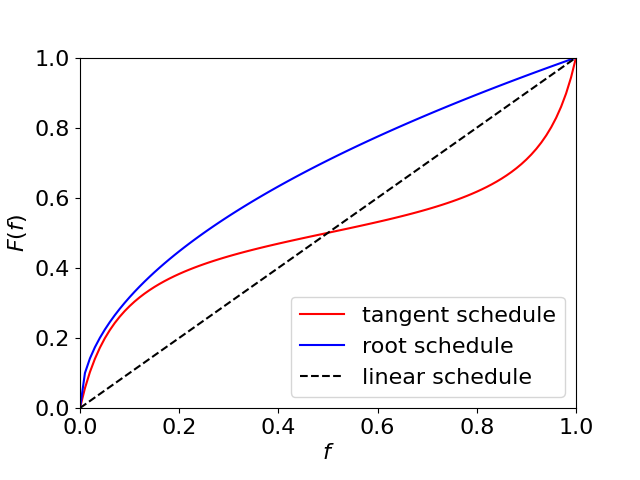}
    \caption{$F(f)$ vs $f$ for the nonlinear schedules}
    \label{fig:nonlinear_ramps}
\end{figure}

Since the minimum energy gap in our chemistry problems
occurs at the end of the evolution \cite{krem2021}, 
our first nonlinear schedule has smaller steps toward the end. We refer to this as the ``root'' schedule. Our second schedule, which we call the ``tangent'' schedule, takes larger steps at the beginning and end, and smaller steps in the middle. 
Our nonlinear schedules modify
Eq.~\ref{e:gamma and beta}
by replacing $f$ on the right-hand sides by a function $F(f)$. 
Specifically, the root schedule is
\begin{equation}
    F(f) = \sqrt{f}
    \label{e:root_f}
\end{equation}
and the tangent schedule is
\begin{equation}
    F(f) = \frac{1}{2\tan(\frac{1}{2c})}\left(\tan(\frac{f-0.5}{c}) + \tan(\frac{1}{2c})\right)
    \label{e:tan_f}
\end{equation}
where $c=0.37$. Fig.~\ref{fig:nonlinear_ramps} shows these choices of $F(f)$ and contrasts them with the linear schedule.

\begin{figure}[H]
    \centering
    \includegraphics[width=2.5 in]{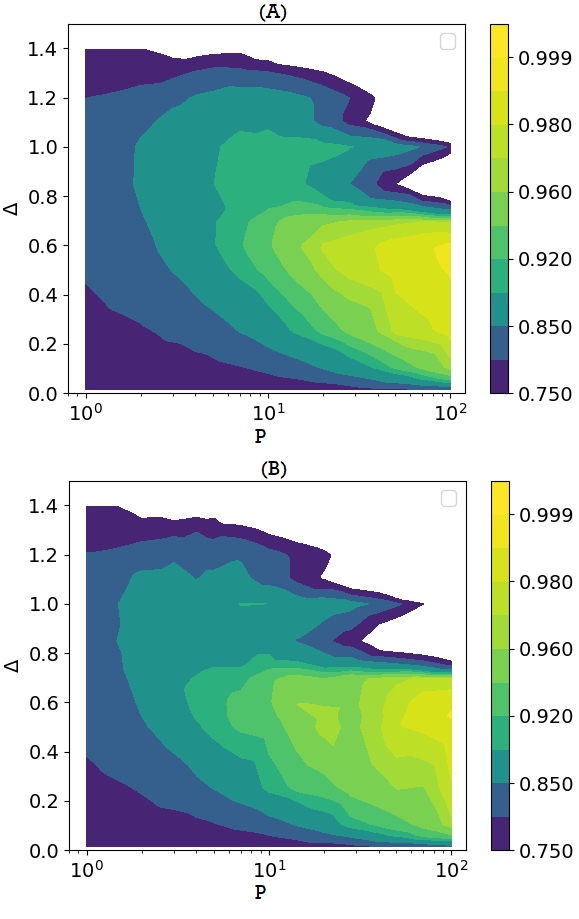}
    \caption{Squared overlap with ground state after completing QAOA as a function of QAOA parameters $\Delta$ and $p$ for P\textsubscript{2}, using (A) root, and (B) tangent schedules.}
    \label{fig:p2_nonlinear}
\end{figure}

\begin{figure}[H]
    \centering
    \includegraphics[width=2.5 in]{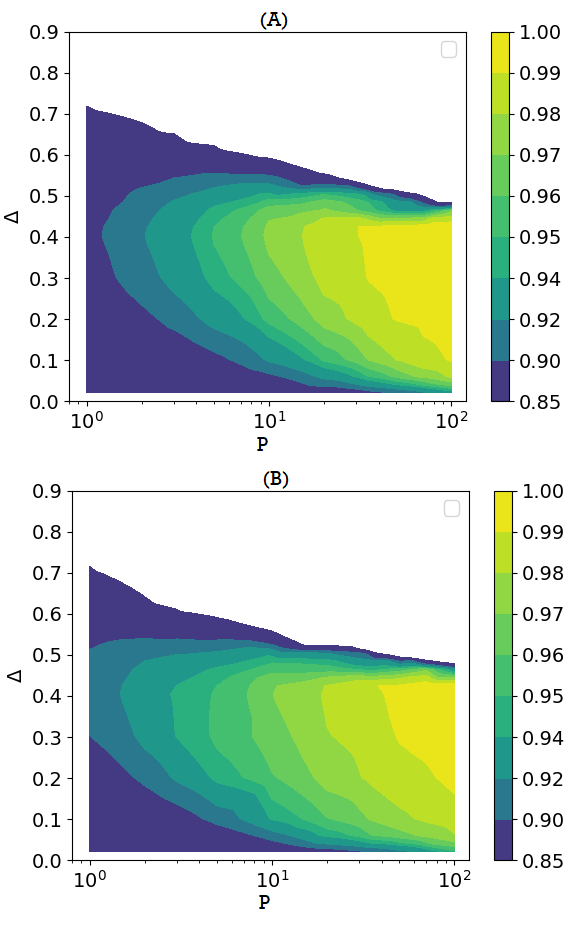}
    \caption{Squared overlap with ground state after completing QAOA as a function of QAOA parameters $\Delta$ and $p$ for CO\textsubscript{2}, using (A) root, and (B) tangent schedules.}
    \label{fig:co2_nonlinear}
\end{figure}

\begin{figure}[H]
    \centering
    \includegraphics[width=2.4 in]{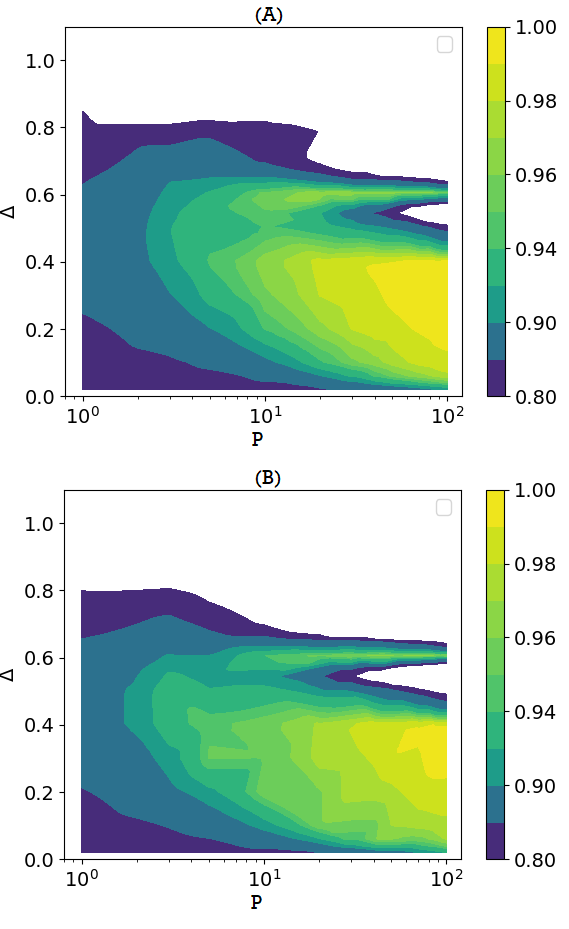}
    \caption{Squared overlap with ground state after completing QAOA as a function of QAOA parameters $\Delta$ and $p$ for Cl\textsubscript{2}, using (A) root, and (B) tangent schedules.}
    \label{fig:cl2_nonlinear}
\end{figure}

\begin{figure}[H]
    \centering
    \includegraphics[width=2.5 in]{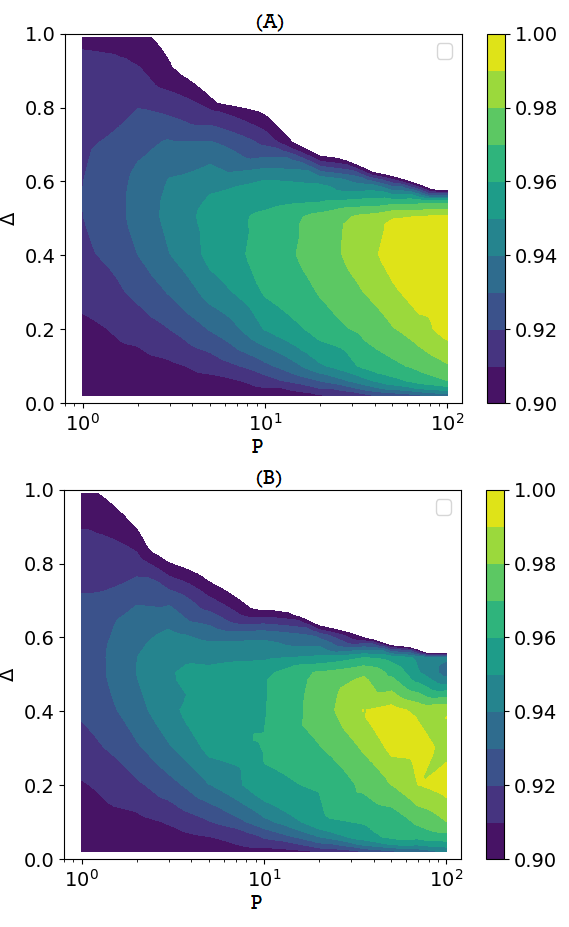}
    \caption{Squared overlap with ground state after completing QAOA as a function of QAOA parameters $\Delta$ and $p$ for CH\textsubscript{2}, using (A) root, and (B) tangent schedules.}
    \label{fig:ch2_nonlinear}
\end{figure}

\vspace{-3mm}
For all four molecules, we find both nonlinear schedules produce the same qualitative features in the QAOA phase diagram as the linear schedule. Moreover, these nonlinear schedules do noticeably worse than the linear ramp, and do comparably as well as each other. The decrease in performance can be seen by the larger $p$ required to reach over $0.99$ squared overlap for small, non-zero $\Delta$. 
To illustrate both of these points, Figs.~\ref{fig:p2_nonlinear} to \ref{fig:ch2_nonlinear} show the root and tangent ramps for the four molecules shown in Fig.~\ref{fig7}. 
\vspace{10mm}
\vspace{-8mm}
\vspace{-8.9mm}
\vspace{-15mm}
\vspace{-6mm}

\section{Satisfiability and Ising spin problems} \label{sec:appB}

A SAT problem is to find an assignment, either true or false, to each of $n$ Boolean variables such that a given logical formula is true. For random 3-SAT, the formula is a conjunction of $m$ clauses. Each clause is a disjunction of a randomly selection of 3 distinct variables, each of which is negated with probability $1/2$. 
This instance has several ground states.

The Ising spin problem is to find the ground state of $n$ spins $s_i$, $i=1,\ldots,n$ each of which is $\pm1$. The fully-connected version illustrated here has Hamiltonian
$$H= \sum_i h_i s_i + \sum_{i<j} J_{i,j} s_i s_j$$
A random instance has the $h_i$ and $J_{i,j}$ selected independently and uniformly at random in the range $-1$ to $1$.

\section{QAOA for Chemistry using XY-mixers} \label{sec:appC}
An alternative QAOA approach  to 
chemistry and materials problems is using mixers related to the XY model, as previously proposed for graph coloring problems~\cite{hadfield2019quantum}. This mixer has the useful property of preserving total particle number, in contrast to the transverse-field mixer originally considered for QAOA.  
This alternative uses the same cost Hamiltonian as in the main text, i.e.,  $H_C=H_e$, 
but differs in the choice of initial state and mixing operator. 

\paragraph{Initial states:}Any linear combinations of strings with Hamming weight $n$ is feasible, corresponding to the $n$-electron sector of the Fock space. Following \cite{hadfield2019quantum}, we select such a feasible state as 
QAOA initial state. For example, we may again use the 
Hartree-Fock state. Alternatively,  
we could use the equal superposition of all $n$-particle basis states in $N$ orbitals,
$\binom{N}{n}^{-1/2}\sum_{x:|x|=n} \ket{x}$; such \textit{Dicke states} can be constructed efficiently
\cite{bartschi2019deterministic}.  Note that here we relax the requirement that the initial state corresponds to the ground state of the mixing operator.

\paragraph{Mixing operator:} 
By careful selection of the mixing Hamiltonian we can ensure the QAOA evolution remains in the feasible subspace~\cite{hadfield2019quantum}. In particular 
the Hamiltonian
$ H_B = \sum_{j<k} X_j X_{k} + Y_j Y_{k}$
which generates the \textit{XY mixer} suffices; see \cite{hadfield2019quantum,wang2020xy} for other variants of this mixer with different resource tradeoffs. 
With this choice 
both the phase and mixing operators are nondiagonal in the computational basis. We emphasize that as the antisymmetry requirement of the problem is enforced by the state encoding (i.e., basis states correspond to Slater determinants), it is not necessary to select a mixer built from simple combinations of Fermionic operators.

\end{document}